\documentclass[preprint,aip]{revtex4-1}
\usepackage{amsmath}%
\usepackage{amsfonts}
\usepackage{amssymb} 
\usepackage{hyperref}
\usepackage{graphicx}
\usepackage{float}
\usepackage{graphics, epsfig, subfig}
\usepackage{epstopdf}
\usepackage{color}
\usepackage{natbib}
\usepackage{esint}

\begin{document}
\title{A quasi-linear analysis of the impurity effect on 
turbulent momentum transport and residual stress}
\author{S.H. Ko}
\email{shko@nfri.re.kr}
\author{Hogun Jhang} 
\author{R. Singh}
\affiliation{National Fusion Research Institute, Daejeon 305-333, 
Republic of Korea} 

\begin{abstract}
We study the impact of 
impurities on 
turbulence driven intrinsic rotation (via residual stress) in 
the context of the quasi-linear theory.
A two-fluid formulation for main and impurity ions
is employed to study
ion temperature gradient modes in sheared slab geometry
modified by the presence of impurities.
An effective form of the parallel Reynolds stress is derived 
in the center of mass frame of a coupled main ion-impurity
system. Analyses show that the contents and the radial profile of 
impurities have a strong influence on the residual stress.
In particular, an impurity profile aligned with that 
of main ions is shown to 
cause a considerable reduction of the residual stress, which 
may lead to the reduction of turbulence driven intrinsic rotation.
\end{abstract}

\maketitle
\newpage{}

\section{Introduction}\label{intro}
%
%
The presence of a certain amount of impurities is,
in some sense, an inevitable
consequence of tokamak plasma operation. 
The contents and the radial profile of the dominant impurity 
species depend on the details of discharge conditions, 
plasma-wall interaction, and the physical process of
impurity transport.
In tokamaks, the material constituting 
plasma facing components (PFCs), for instance,
carbon in  DIII-D\cite{DIIID} and KSTAR\cite{KSTAR},
or tungsten in ASDEX-U\cite{AUG}, JET\cite{JET} and ITER\cite{ITER_impW},
is usually observed as a dominant impurity species.
In addition to impurities originating from the surrounding PFCs,
a considerable amount of helium ash 
will be present in ITER or 
fusion reactors as a result of fusion reaction\cite{ITER_impHe,Polevoi}.

%
Studies of impurity transport in tokamaks have usually focused on the determination
of impurity profiles (e.g. the impurity peaking factor) from either 
neoclassical\cite{Hirsh-Sig,NeoKim,Neo2} or
turbulent mechanisms\cite{Garbet,AP}. The main goal of this 
endeavour is to predict
the degree of impurity accumulation for a given set of 
discharge parameters to avoid deleterious effect from impurity accumulation.
Turbulent transport of impurities, which has been a subject of
an extensive study in fusion plasma physics for decades \cite{Garbet,AP,Coppi,Dong1,Dong2,Dong,FW} 
is rekindling interest in this regard
since recent experimental evidence has shown the importance of
the turbulent transport process in determining the core impurity 
profile\cite{Angioni-JET}.

When impurity concentration is sufficiently low, one
may treat the impurities as passive reactants to background turbulence.
In this case, the so-called trace approximation has been used to study
turbulent impurity transport\cite{AP,Nordman}. 
If impurity concentration is considerable, however, 
they start to affect 
microinstability\cite{Coppi,Dong1,Dong2,Dong} and
the trace approximation becomes invalid, as pointed out by F\"{u}l\"{o}p
and Nordman\cite{FN}. Most notably, the alignment
of a impurity profile with that of main ions has been known to yield
a significant change of stability of 
the ion temperature gradient (ITG) mode, sometimes
leading to the destabilization of an independent impurity 
drift mode\cite{Tang,Dong}

Our interest in this paper lies in 
the study of the impact of impurities on 
turbulent momentum transport, in particular, the generation of 
intrinsic rotation via residual stress\cite{RS1,RS2}. 
We note that few results are available in this aspect, 
given a lot of published papers on turbulent impurity transport in tokamaks.
Intrinsic rotation in magnetic fusion plasmas
is thought to be generated through
the conversion of radial inhomogeneity into $\left< k_{||} \right>$ (=
spectrally averaged wavenumber in parallel direction) asymmetry via residual stress. 
This conversion process
requires some symmetry breaking mechanism\cite{Gurcan1,Gurcan2,McDevitt,Singh,Ram}.
An early paper calculated the particle
flux in the presence of combined ITG, impurity and parallel velocity gradients, 
but not addressed the turbulent residual stress giving rise to 
intrinsic rotation\cite{Dong}.

The purpose of this study is to provide a simple physics insight into which how 
turbulence driven residual stress is related to the characteristics of
impurities. Intuitively, one may expect that the change of stability and
characteristics of an unstable mode structure in the presence of a considerable amount
of impurities will alter the turbulent residual stress. This effect
may change the amount of intrinsic torque anticipated in toakamak
experiments. Since intrinsic rotation is envisioned to provide necessary plasma rotation
in reactor-relevant tokamak experiments,
including ITER, because of the limited capability of
driving the necessary torque in a reactor by neutral beam injection, it is of importance
to conduct a sufficiently detailed study of this effect. 

To address the problem mentioned above, we employ a set two fluid 
equations for main and impurity ions. By treating impurities
as a separate species satisfying impurity fluid equations, we can
incorporate the effects of impurities on the characteristics of
the ITG mode self-consistently. 
For simplicity, we adopt the sheared
slab geometry. By avoiding the complication in analysis 
due to the sophisticated toroidal geometry, we focus on the basic physics 
of the impurity effect on residual stress in the presence of
impurity-modified ITG turbulence. 
Analyses show that the relative importance
of residual stress to the diffusive momentum flux 
is reduced when the gradient of impurity and main ions has
the same sign. This indicates
a possible decrease of intrinsic rotation
when a considerable amount of core-peaked impurity is present.


The remainder of the paper is organized as follows. In Sec. II,
we describe the basic formulation. A derivation is given of a linear dispersion
relation and an eigenmode equation
for ITG modes using two fluid equations for ions and impurities.
In Sec.~III, we perform a numerical analysis to
evaluate eigenmode structures of
unstable modes. A detailed study of 
the change of eigenmode characteristics, such as
the variation of the mode shift off the rational surface, 
is made in this section. Section~IV is devoted to 
the quasi-linear calculations
of momentum flux.
After introducing some underlying assumptions made in this study,
we calculate turbulent parallel Reynolds stress (both diffusive and
residual stresses) in the context of the quasi-linear analysis. 
A particular emphasis is placed on the evaluation of the ratio
of residual stress to diffusive one, which ameliorates the limitation
of the quasi-linear theory. 
Possible implications of the results will
be discussed in this section. We conclude this paper in Sec.~V
with a brief summary of main results and some discussions.
%

\section{Formulation}\label{form}

We begin with a set of two fluid equations for main and
impurity ions consisting of the conservation of density ($n_j$), parallel
momentum density $(m_j n_j v_{\parallel j})$, and pressure $P_j$,
\begin{eqnarray}
&&\frac{\partial n_j}{\partial t}+\nabla \cdot(n_j{\mathbf{v}}_j)=0,\nonumber\\
&&m_j n_j\frac{d{\mathbf{v}}_{\parallel j}}{dt}+\nabla_{\parallel}P_j+n_je_j\nabla_{\parallel}\phi=0,\nonumber\\
&&\frac{dP_j}{dt}+\Gamma P_j\nabla\cdot{\mathbf{v}}_{\parallel}=0,
\label{fluids}
\end{eqnarray}
where the subscript $j=i$ ($Z$) denotes 
the main (impurity) ions, $\phi$ is the electrostatic potential fluctuation,
and $\Gamma$ is the adiabatic index. 
The fluid velocity is decomposed into the parallel and
the perpendicular components, 
$\mathbf{v}_j=\mathbf{v}_{\parallel j}+\mathbf{v}_{\perp j}$ with
$\mathbf{v}_{\perp j}=\mathbf{v}_{Ej}+\mathbf{v}_{*j}+\mathbf{v}_{pj}$, where 
$\mathbf{v}_{Ej}$, $\mathbf{v}_{*j}$, and $\mathbf{v}_{pj}$ 
are the $E\times B$, the diamagnetic, and 
the polarization drift of the species $j$, respectively. 
We employ the sheared slab geometry with a
background magnetic field in the vicinity of a reference
surface ($r_0$),
$\vec{B}_0=B_0 \left( \hat{z}+{x}/{L_s}\hat{y} \right)$. Here,
$L_s = qR/rq^\prime$ is the magnetic shear scale length
with $q$ the safety factor, $R$ ($r$) the major (minor) radius, and
the prime denotes a derivative with respect to $x=r-r_0$.

Linearization of Eq.~(1) can be done straightforwardly giving rise to
%
\begin{subequations}
\begin{align}
&d_t^E(n_i-\nabla^2_{\perp}\phi)+L_{ei}(1+K_i\nabla^2_{\perp})\nabla_y \phi+\nabla_{\parallel}v_{\parallel i}=0,\\
&d_t^Ev_{\parallel i}-\hat{V}_{0\parallel i}' \nabla_y\phi
+\nabla_{\parallel}\phi+\tau_i\nabla_{\parallel}p_i =0, \\
&d_t^Ep_i+L_{ei}(1+\eta_i)\nabla_y\phi+\Gamma\nabla_{\parallel}v_{\parallel i}=0
,\\
&d_t^E(n_Z-\frac{\mu}{Z}\nabla^2_{\perp}\phi)+L_{eZ}(1+\frac{\mu}{Z^2}K_Z\nabla^2_{\perp})\nabla_y \phi+\nabla_{\parallel}v_{\parallel Z}=0
,\\
&d_t^Ev_{\parallel Z}-\hat{V}_{0\parallel Z}' \nabla_y \phi+\frac{Z}{\mu}\nabla_{\parallel}\phi+\frac{\tau_Z}{\mu}\nabla_{\parallel}p_Z=0,\\
&d_t^Ep_Z+L_{eZ}(1+\eta_Z)\nabla_y\phi+\Gamma\nabla_{\parallel}v_{\parallel Z}=0,
\label{two_fluid}
\end{align}
\end{subequations}
where $d_t^E=(\partial_t + x \hat{V}_{E0}^\prime \nabla_y)$ with 
$\hat{V}_{E0}^\prime$ the normalized equilibrium $E\times B$ flow shear. 
Various quantities in Eq.~(2) are normalized as 
$x=x/\rho_s$, $y=y/\rho_s$, $z=z/L_{ne}$, $t=t/(L_{n_e}/c_s)$, $\phi=(e\phi/T_e) (L_{n_e}/\rho_s)$, 
$n_i=(n_{i1}/n_{i0}) (L_{n_e}/\rho_s)$, $n_Z=(n_{Z1}/n_{Z0}) (L_{n_e}/\rho_s)$, $v_{i||}=(v_{i\parallel1}/c_s) (L_{n_e}/\rho_s)$, $v_{Z||}=(v_{Z\parallel1}/c_s) (L_{n_e}/\rho_s)$,
$\hat{V}_{E0}'=(L_{n_e}/c_s) (dV_{E0}/dx)$, $\hat{V}_{0\parallel i}'=(L_{n_e}/c_s) (dV_{i0\parallel}/dr)$, $\hat{V}_{0\parallel Z}'=(L_{n_e}/c_s) (dV_{Z0\parallel}/dr)$, 
$p_i=(P_{i1}/P_{i0}) (L_{n_e}/\rho_s)$, $p_Z=(P_{Z1}/P_{Z0}) (L_{n_e}/\rho_s)$
with
$L_{ne}^{-1}=-\nabla n_0/n_0$ the equilibrium density scale length, $c_s=\sqrt{T_{e0}/m_i}$ ($T_{e0}$: equilibrium electron temperature),
and $\rho_s=c_s/\Omega_i$ the ion-acoustic Larmor radius. 
Also, various dimensionless parameters are defined as
$\mu=m_Z/m_i$, 
$\eta_i = {L_n}/{L_{Ti}}$ with $L_{Ti}^{-1}=-\nabla T_{0i}/T_{0i}$
the equilibrium ion temperature scale length,
$\eta_Z = L_{nZ}/L_{TZ}$,
$K_i = \tau_{i}(1+\eta_i)$ with $\tau_i=T_{i0}/T_{e0}$, 
$K_Z = \tau_{Z}(1+\eta_Z)$ with $\tau_Z=T_{Z0}/T_{e0}$, 
$L_{ei}=L_{n_e}/L_{n_i}$, and $L_{eZ}=L_{n_e}/L_{n_Z}$.

We consider perturbations of the form 
$f=f_k(x)\exp\left( ik_y y+ik_zz-i\omega t \right)$, where $\omega$ is normalized to $c_s/L_n$. 
Then, one can calculate ion and impurity density fluctuations from
Eqs.~(2a) to (2f),
\begin{eqnarray*}
n_{ik}=-\frac{1}{\hat{\omega}}\left[\left(\hat{\omega}+k_y L_{ei}K_i \right)
k_{\perp}^2 - k_yL_{ei} - \frac{k_\parallel \hat{\omega}}
{\hat{\omega}^2-k_{\parallel}^2\tau_i\Gamma}\left({k_{\parallel}-k_y\hat{V}_{0\parallel i}'}+\frac{k_yk_{\parallel}L_{ei}K_i}{\hat{\omega}}\right)\right]\phi_k,
\label{ni_fluc}
\end{eqnarray*}
and
\begin{eqnarray*}
n_{Zk}=-\frac{1}{\hat{\omega}}\left[\frac{\mu}{Z}\left(\hat{\omega} + 
\frac{k_y L_{eZ}K_Z}{Z} \right) k_{\perp}^2 - {k_yL_{eZ}}
- \frac{k_\parallel \hat{\omega}}
{\hat{\omega}^2-k_{\parallel}^2 ({\tau_Z}/{\mu})\Gamma}  
\left({k_{\parallel}-k_y\hat{V}_{0\parallel Z}'}
+\frac{k_yk_{\parallel}L_{eZ}K_Z}{\hat{\omega}}\right) \right]\phi_k,
 \label{nZ_fluc}
\end{eqnarray*}
\noindent where $\hat{\omega}=\omega-k_yx\hat{V}_{E0}'$ is the normalized
frequency of the fluctuation accounting for the Doppler shift. 

Ions, impurities, and electrons are coupled through the quasi-neutrality 
condition, $n_{ek}=(1-f_Z)n_{ik}+f_Zn_{Zk}$ where
$f_Z={Zn_{Z0}}/{n_{e0}}$ is the fraction of impurity under consideration.
Assuming the adiabatic electron response, $n_{ek}=\phi_k$,
one can derive, after a little algebra, an eigenvalue
equation in $\phi_k$,
\begin{equation}
\frac{d^2\phi_k}{d x^2}+U(k_{\parallel},k_y,\hat{\omega})\phi_k=0,
\label{general_eigen}
\end{equation}
where
\begin{eqnarray}
U(k_{\parallel},k_y,\hat{\omega})&=&-k_y^2+\frac{k_y-\hat{\omega}}{\alpha\hat{\omega}+k_yA} + \frac{(1-f_Z)\hat{\omega}}{(\alpha\hat{\omega}+k_yA)(\hat{\omega}^2-\tau_i k_{\parallel}^2\Gamma)}\left(k_{\parallel}^2-k_yk_{\parallel}
\hat{V}_{0\parallel i}'+\frac{k_yk_{\parallel}^2L_{ei}K_i}{\hat{\omega}}\right)\nonumber\\
&&+\frac{f_Z\hat{\omega}}{(\alpha\hat{\omega}+k_yA)\left[\hat{\omega}^2-({\tau_Z}/{\mu}) k_{\parallel}^2\Gamma \right]}\left(\frac{Z}{\mu}k_{\parallel}^2-k_yk_{\parallel}\hat{V}_{0\parallel Z}'+\frac{k_yk_{\parallel}^2L_{eZ}K_Z}{\mu\hat{\omega}}\right)\label{U_org}
\end{eqnarray}
with the definitions $\alpha= 1-f_Z+({\mu}/{Z})f_Z$
and $A=(1-f_Z)L_{ei}K_i+f_Z ({\mu}/{Z^2}) L_{eZ} K_Z$. 
Equation~(\ref{U_org}) represents a complete potential function for the
ITG eigenmode modified by the presence of impurities whose fraction
is $f_Z$. Following Singh {\it et. al.,}\cite{Ram}, 
we assume that the mode frequency is much higher than
the equilibrium shearing rate and the ion acoustic
wave frequency, {\it i.e.,} 
$\omega^2\gg k_\parallel^2 \tau_i \Gamma$ and $\omega \gg k_y 
\hat{V}_{E0}^\prime$. We further assume that the equilibrium
ion and impurity velocity profiles are equal, 
$\hat{V}_{0\parallel i}=\hat{V}_{0\parallel Z}=\hat{V}_{0\parallel}$.
Under these assumptions, Eq.~(\ref{U_org}) is reduced to
\begin{equation}
U(k_y,\omega,x)=A_0+A_1x+A_2x^2,
\label{U}
\end{equation}
where
\begin{eqnarray}
&&A_0=-k_y^2+\frac{k_y-\omega}{\alpha\omega+k_yA},\nonumber\\
&&A_1=\frac{k_y}{\alpha\omega+k_yA}\left(\hat{V}_{E0}^\prime-\frac{k_ys}{\omega}\hat{V}_{\parallel i}'\right),\nonumber\\
&&A_2=\frac{\beta\omega+k_yB}{\alpha\omega+k_yA}\frac{k_y^2s^2}{\omega^2},\nonumber
\end{eqnarray}
with the definitions $\beta=1-f_Z+({Z}/{\mu})f_Z$ and $B= (1-f_Z)L_{ei}K_i+({f_Z}/{\mu})L_{eZ}K_Z$.  
Equation~(5) with the
coefficients $A_0, A_1, A_2$ is a simplified form of the impurity-modified
potential function for the ITG mode
with the frequency $\hat{\omega}$ and the
mode number $k_y$. The impurity effect
on mode characteristics are now contained in the parameters
$\alpha$, $A$ and $B$ in this formulation.

Solutions to Eq.~(\ref{general_eigen}) are well-known Hermite polynomials. 
To make an analytical progress, we consider the most dominant zeroth 
order Hermite polynomial. Then, the eigenfunction becomes
\begin{eqnarray}
\phi_{k_y}&&=\phi_0\exp\left[-\frac{1}{2}i\sqrt{A_3}\left(x+\frac{A_2}{A_3}\right)^2\right]\nonumber\\
&&=\Phi_{0k_y}\exp\left[-\frac{1}{2}\left(\frac{x-X_0}{\Delta_{k_y}}\right)^2\right],
\label{eigen_fun}
\end{eqnarray}
where
\begin{equation}
X_0=-\left(Re\frac{A_2}{2A_3}+\frac{Re\sqrt{A_3}}{Im\sqrt{A_3}}Im\frac{A_2}{2A_3}\right)
\label{eq_mode_shift}
\end{equation}
is the shift of the mode from the mode rational surface,
\begin{equation}
\Delta^{-2}_{k_y}=-Im\sqrt{A_3}
\label{width}
\end{equation}
is the width of the mode, and
\begin{eqnarray}
\Phi_{0k_y}&&=\Phi_0 (A_1,A_2,A_3) \exp\left[-\frac{i}{2}Re\sqrt{A_3}\left(x+Re\frac{A_2}{2A_3}-\frac{Im\sqrt{A_3}}{Re\sqrt{A_3}}Im\frac{A_2}{2A_3}\right)^2\right].
\end{eqnarray}
We plug Eq.~(\ref{eigen_fun}) into Eq.~(\ref{general_eigen}) and evaluate the
resulting relation,
\begin{eqnarray}
\left(A_1-\frac{A_2^2}{4A_3}\right)/\sqrt{-A_3}=1,\nonumber
\label{dispersion_gen1}
\end{eqnarray}
which converts into the following dispersion relation,
\begin{eqnarray}
-k_y^2+\frac{k_y-\omega}{\alpha\omega+k_yA}-\frac{k_y^2}{4D(\alpha\omega+k_yA)^2}\left(\frac{\omega}{k_ys}\hat{V}_{E0}'-\hat{V}_{0\parallel}'\right)^2=i\sqrt{D}\frac{k_ys}{\omega},
\label{dispersion_gen2}
\end{eqnarray}
\noindent where $D=(\beta\omega+k_yB)/(\alpha\omega+k_yA)$.
One can rearrange Eq.~(\ref{dispersion_gen2}) to obtain 
\begin{equation}
\omega^2(1+\alpha k_y^2)+k_y\omega(-1+k_y^2A+i\sqrt{D}s\alpha)+i\sqrt{D}k_y^2sA=-\frac{k_y^2\omega}{4D(\alpha\omega+k_yA)}\left(\frac{\omega}{k_ys}\hat{V}_{E0}'-\hat{V}_{0\parallel}'\right)^2.
\label{dispersion_rel}
\end{equation}
It is an easy task to show that Eq.~(\ref{dispersion_rel}) recovers the
dispersion relation without impurities\cite{Singh}
when $f_Z=0$, $\alpha=1$, $A=K_i$ and $D=1$.
%
%
\begin{figure}
\begin{center}
\includegraphics[scale=0.8]{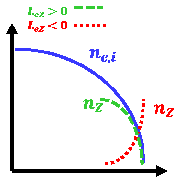}
\caption{A schematic diagram illustrating the
wall-peaked ($L_{eZ}<0$, red dotted) and
the core-peaked ($L_{eZ}>0$, green dashed) impurity 
profiles. Solid blue line represents an equilibrium
plasma density profile.}
\label{schematic_diag}
\end{center}
\end{figure}
%
%
%

Equation~(\ref{dispersion_rel}) is the desired equation to be 
analysed in this paper. It has been known that
the radial profile of an impurity has a significant influence on
ITG stability\cite{Coppi,Tang,Dong1,Dong2}.
In the case of a wall-peaked impurity profile 
schematically shown in
Fig.~1 (red dotted line), the ITG mode is destabilized while
the core-peaked one (green dashed line in Fig.~1), stabilizes it. 
Therefore, before performing a detailed numerical
analysis, it is instructive to examine if Eq.~(\ref{dispersion_rel})
recovers these known results. 
To make the analysis simple, we first consider
Eq.~(\ref{dispersion_rel}) in the slow mode 
({\it i.e.,} the long wavelength) limit\cite{Singh,Balescu}
which is characterized by the relation $|k_y^2A|\ll 1$. 
We further assume that $\hat{V}_{E0}'=\hat{V}_{0\parallel}'=0$, and $s\ll 1$. 
Then, one obtains a purely growing mode with the growth rate,  
\begin{eqnarray}
\omega_{0i}\sim ik_ysA&&=ik_ys\left[(1-f_Z)L_{ei}K_i+f_Z\frac{\mu}{Z^2}L_{eZ}K_Z\right]\nonumber \\
&&=ik_ys\left[K_i-f_ZL_{eZ}\left(K_i-\frac{\mu}{Z^2}K_Z\right)\right],
\label{limit_freq}
\end{eqnarray}
\noindent where the relation $L_{ei}=\left({1-f_ZL_{eZ}}\right)/(1-f_Z)$
is used from the quasi-neutrality condition. 
From Eq.~(\ref{limit_freq}), one can see that the mode becomes 
stabilized when $L_{eZ}>0$ ({\it i.e.,} 
the core-peaked case), while it is destabilized when $L_{eZ}<0$ ({\it i.e.,} the wall-peaked case). 
In slab ITG modes, 
this can be interpreted by an effective reduction or enhancement of 
negative compressibility due to impurity inertia.
We now consider the other limit 
where the relation $|1-k_y^2A|\le s\ll1$ holds, {\it i.e.,}
the fast mode limit. In this limit, the fastest growing mode
occurs at $k_y^2=A^{-1}$, giving rise to the dispersion relation,
\begin{equation}
\omega=(-1+i)\sqrt{\frac{s}{2(1+\alpha/A)}}.\nonumber
\end{equation}
Again, the mode becomes stabilized (destabilized) when 
$L_{eZ}>0$ ($L_{eZ}<0$) due to the decrease (increase) of $A=(1-f_Z)L_{ei}K_i+(\mu/Z^2)f_Z L_{eZ}K_Z$.

\section{Eigenmode analysis}\label{disp}

\begin{figure}
\centering
\includegraphics[scale=0.6]{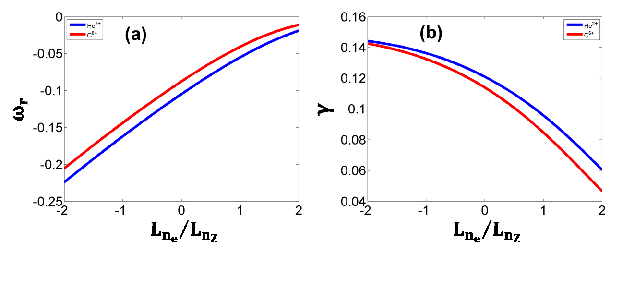}
\caption{The real frequency ($\omega_r$) (a) and the growth rate ($\gamma$)
of the most unstable eigenmode corresponding to $k_y=0.5$
as a function of the ratio between 
the equilibrium scale length of the electron 
($L_{ne}=-n_{0}/\nabla n_0$) to the impurity ($L_{nZ}=-n_{Z0}/\nabla n_{Z0}$)
density.
$\rm{He}^{2+}$ (blue) and $\rm{C}^{6+}$ (red) 
are considered as representative impurities in tokamaks.
Deuterium is assumed to be the main ion species. 
Plasma parameters being used are $\tau_i=\tau_Z=1$, $f_Z=0.3$,
$\eta_i=2$, $\hat{V}_{E0}'=-0.01$, and 
$\hat{V}_{0\parallel}'=-0.1$.} 
\label{frequency_Lez}
\end{figure}

Figure~\ref{frequency_Lez}(a) and 2(b) show the real frequency ($\omega_r$) 
and the growth rate ($\gamma$), respectively,
as a function of $L_{eZ}=L_{n_e}/L_{n_Z}$ for 
two representative impurity ions in tokamaks, ${\rm {He}^{2+}}$ 
in blue and $\rm{C}^{6+}$ in red. 
To produce Fig.~\ref{frequency_Lez} we 
solved Eq.~(\ref{dispersion_rel}) numerically using the parameters,
$\tau_i=\tau_Z=1$, $f_z=0.3$, $k_y=0.5$,
$\eta_i=2$, $\hat{V}_{E0}'=-0.01$, and $\hat{V}_{0\parallel}'=-0.1$.
The wavenumber $k_y=0.5$ corresponds to the most unstable mode giving
rise to the maximum growth rate in this case. 
The equilibrium $E\times B$ shear $\hat{V}_{E0}'=-0.01$ roughly corresponds to
$E_r \approx -5~kV$ if we assume $B_T = 1~\rm{Tesla}$ and $T_e \approx 3~keV$.
This is approximately $\sim 1/4$ of the radial electric field well
in typical H-mode discharges\cite{Er-DIII,Er-JET}.
Deuterium is used as the main ion 
species in these calculations and throughout
the paper. Figure~2 basically confirms the analytic 
estimation of the impact of $L_{eZ}$ on ITG stability presented in Sec.~II: stabilization (destabilization) of the ITG mode
by a wall-peaked (core-peaked) impurity profile.
The lighter impurity species (${\rm He}^{2+}$) shows a little higher real frequency
and growth rate compared to a heavier one, but the deviation
is insignificant as long as the $f_Z$ value is fixed. 
%
%
%
%

\begin{figure}
\includegraphics[scale=0.6]{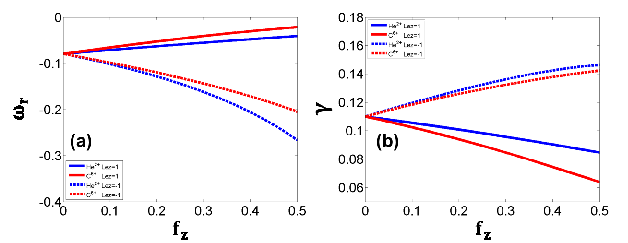}
\caption{The real frequency ($\omega_r$) (a) and the growth rate ($\gamma$)
of the most unstable eigenmode as a function of impurity contents ($f_Z$).
Solid and dotted lines represent when $L_{eZ}=1$ and $-1$, respectively.
$\rm{He}^{2+}$ (blue) and $\rm{C}^{6+}$ (green) 
are considered as representative impurities. 
The same plasma parameters (except for $f_Z$) as those 
in Fig.~2 have been used.}
\label{frequency_fz}
\end{figure}

The impact of impurity fraction, $f_Z$, on the eigenfrequncy is shown
in Figs.~\ref{frequency_fz}(a) and 3(b). In these figures, we consider
two different impurity profiles: $L_{eZ}=1$ (solid) and  $L_{eZ}=-1$ (dotted).
Other parameters to produce Fig.~\ref{frequency_fz} are same as those of
Fig.~\ref{frequency_Lez}. The stabilization ($L_{eZ}=1$) or 
destabilization ($L_{eZ}=-1$) of the ITG mode by impurities
becomes stronger as $f_Z$ increases.  
An interesting observation is that there is asymmetry in
the stabilization and destabilization by impurities
for a fixed $f_Z$ value, as can
be seen in Fig.~\ref{frequency_fz}(b). Thus, a heavy 
impurity with a core-peaked profile has a stronger stabilizing
influence compared to the lighter impurity, 
while the destabilization effect due to it with a wall-peaked profile 
is weaker than that of the latter.  
%
%

\begin{figure}
\includegraphics[scale=0.6]{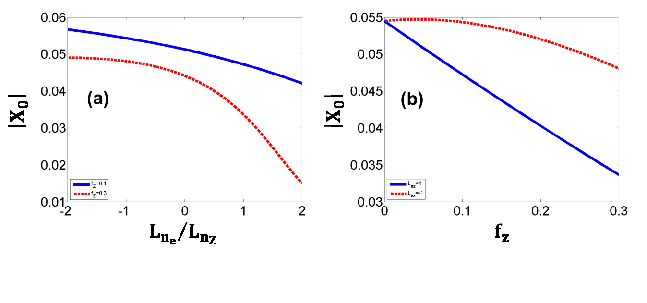}
\caption{\label{mode_shift} 
The shift ($X_0$) of the most unstable ITG mode ($k_y=0.5$)
off the rational surface 
as a function of $L_{eZ}=L_{n_e}/L_{n_Z}$ (a) and $f_Z$ (b). 
Blue solid and red dotted lines  
in (a) represent when $f_Z=0.1$ and 0.3, respectively. 
In (b), blue solid and red dotted lines  
show when $L_{eZ}=1$ and $-1$, respectively.
$\rm {C}^{6+}$ impurity is used with 
plasma parameters $\tau_i=\tau_Z=1$,
$\eta_i=2$, $\hat{V}_{E0}'=-0.01$, and 
$\hat{V}_{0\parallel}'=-0.1$.}
\end{figure}

Of particular interest in this paper is the strength of 
the mode shift ($X_0$) given
in Eq.~(\ref{eq_mode_shift}). This is because $X_0$ driven by
the $E \times B$ shear is an essential quantity in the calculation of 
residual stress\cite{Gurcan1,Gurcan2,Singh,Ram}, hence the intrinsic
rotation, in the quasi-linear theory.  
%
Figure~\ref{mode_shift}(a) shows $|X_0|$ vs. $L_{n_e}/L_{n_Z}$ for $\rm{C}^{6+}$
when $f_z=0.1$ (solid) and 0.3 (dotted). 
In general, $|X_0|$ for a given value of 
$\hat{V}_{E0}^\prime$ decreases as the
local impurity profile changes from wall- to core-peaked one, 
{\it regardless of $f_Z$}. 
Thus, one may expect that the ITG-driven
residual stress will increase as the impurity profile changes from 
the wall to core-peaked one. 
The trend of $|X_0|$ reduction shows a remarkable difference 
depending on the impurity contents.
When $f_Z=0.1$, $|X_0|$ shows an almost linear behaviour
in the whole range of  conceivable $L_{n_e}/L_{n_Z}$ values, as
can be seen in Fig.~4(a).
When  $f_Z=0.3$, however, $|X_0|$ does not change much when negative
$L_{n_e}/L_{n_Z} < 0$, 
while it starts to decrease very rapidly as soon as 
$L_{n_e}/L_{n_Z}$ becomes positive. 
For example, the reduction in $|X_0|$ as $L_{n_e}/L_{n_Z}$ changes
from $0$ to $1$ is $\sim 25\%$ and $\sim 9\%$ when $f_Z=0.3$ and
$0.1$, respectively.

Figure~\ref{mode_shift}(b) shows the effect of $f_Z$ on $|X_0|$ 
when $L_{n_e}/L_{n_Z}=1$~(solid) and $-1$~(dotted), respectively. 
Unlike the growth rate, $|X_0|$ always decreases as $f_Z$ increases,
regardless of the impurity profile shape.
The decrease of $X_0$ 
can be seen more clearly in the core-peaked impurity profile 
which exhibits a significant reduction of
$|X_0|$ when $f_Z$ increases. The 
impurity effect on $X_0$ 
is not considerable for a wall-peaked profile with
a moderate dilution of a plasma, {\it{i.e.,}}, when $f_Z < 0.2$. 
A core-peaked profile, however, will result in large 
reduction of $X_0$, which
eventually leads to a significant decrease of residual stress. 
This is a subject for the next section.

\section{Momentum Flux and Residual Stress}\label{mom_flux}
In this section, we calculate the parallel Reynolds stress
in the presence of a small fraction of impurity ions. 
To begin, we add parallel force balance equations for ions and impurities, Eqs.~(2b) 
and (2e), respectively, and
take an ensemble average of the resulting equation to obtain 
\begin{eqnarray}
\frac{\partial \left< V_\parallel \right>}{\partial t} =
- \frac{\partial}{\partial r} 
\left[\Pi^{(i)}_{r||}+\frac{m_Z n_{Z0}}{m_in_{i0}}\Pi^{Z}_{r||} \right] 
+ \frac{1}{m_i n_{i0}}
\left[ \left<\tilde{n}_i\nabla_{\parallel}\tilde{T}_i \right>
+ \left<\tilde{n}_Z\nabla_{\parallel}\tilde{T}_Z \right]
\right>.
\label{Effective_V}
\end{eqnarray}
\noindent In Eq.~(\ref{Effective_V}), 
$\left< A \right>$ represents the ensemble average of a 
quantity $A$. In this study, we 
define an {\it {effective}} 
fluid velocity as $\left< V_\parallel \right> = \left<V_{\parallel i}
\right> +\left[(m_Z n_{Z0})/(m_in_{i0})\right]\left<V_{\parallel Z} 
\right>$, which is the center of mass (CM) velocity in an equilibrium
state of a combined ion-impurity system. 
$\Pi^{(i)}_{r\parallel}$ and
$\Pi^{(Z)}_{r\parallel}$ are
the contribution to the total parallel Reynolds 
stress from main ions and impurities,
respectively. The last two terms in Eq.~(\ref{Effective_V})
represent the parallel turbulence acceleration 
whose characteristics have been studied elsewhere\cite{Wang,Garbet-TA}. 

In this study, we do not take the turbulence acceleration effect into
account, focusing ourselves on 
the residual stress due to the symmetry breaking by
the $E\times B$ shear.
Also, we assume that
the equilibrium flow velocities of main ions and impurities under 
consideration is equal,  
${V}_{0\parallel i}={V}_{0\parallel Z} \equiv V_{0||}$.
This assumption is strictly valid at the edge region of a tokamak plasma 
with sufficiently high collisionality. 
We keep this assumption in this study by taking notice of the
validity of our analysis.  
Then, Eq.~(\ref{Effective_V}) is simplified into the form, 
\begin{equation}
\frac{\partial \left< V_\parallel \right>}{\partial t}=-\nabla_r\Pi^{\rm {eff}}_{r\parallel},
\label{tot_momentum}
\end{equation}
\noindent \noindent where $\Pi^{\rm {eff}}_{r\parallel}$ 
is the effective parallel Reynolds stress of the coupled
ion-impurity system.
One can write  $\Pi^{\rm {eff}}_{r\parallel}$ in a canonical form in the
quasilinar theory, 
\begin{equation}
\Pi_{r\parallel}^{\rm eff}=-\chi_{\phi}^{\rm {eff}}\frac{\partial V_{\parallel 0}}{\partial r}+\Pi_{res}^{\rm {eff}}
\label{tot_Reynolds}
\end{equation}
\noindent with $\chi_{\phi}^{\rm {eff}}$ 
and $\Pi_{res}^{\rm {eff}}$ the effective momentum 
diffusivity and the residual stress defined at the CM frame, 
respectively.
Then, it is easy to derive expressions for $\chi_{\phi}^{\rm {eff}}$
and $\Pi_{res}^{\rm {eff}}$ to obtain 
\begin{equation}
\chi_{\phi}^{\rm {eff}}=\frac{1}{1+\hat{M}}\chi^i_{\phi}+\frac{\hat{M}}{1+\hat{M}}\chi^Z_{\phi},
\label{tot_diffusivity}
\end{equation}
\begin{equation}
\Pi_{res}^{\rm {eff}}=\frac{1}{1+\hat{M}}\Pi^i_{res}+\frac{\hat{M}}{1+\hat{M}}\Pi^Z_{res},
\label{tot_residual}
\end{equation}
\noindent where $\hat M = m_Zn_{Z0}/m_in_{i0}$, 
$\chi^i_{\phi}$ ($\chi^Z_{\phi}$) is the turbulent ion (impurity) 
momentum diffusivity, and 
$\Pi^i_{res}$ ($\Pi^Z_{res}$) is the residual stress whose radial gradient
brings about the intrinsic torque.
The physical origin for the appearance of
$\hat{M}$ is obvious. Since we assume an equal
flow velocity for main ions and impurities, the turbulence
driven intrinsic torque is distributed between 
them according to their inertia. This effect is
represented by $\hat{M}$ in Eqs.~(\ref{tot_diffusivity}) and (\ref{tot_residual}).

To complete the derivation, we need to evaluate the momentum
diffusivities and parallel residual stress terms. 
It is straightforward to calculate them to obtain
\begin{eqnarray}
\Pi^i_{r||}=\langle\delta v_{E_r}\delta v_{\parallel i}\rangle &=&\left(\frac{c_s\rho_s}{L_{n_e}}\right)^2\sum_{\vec{k}} \frac{\gamma k_y^2}{|\omega|^2}\left[-\hat{V}_{0\parallel i}'+\left< \frac{k_{\parallel}}{k_y} \right>\left(1+\frac{2k_y\omega_rL_{ei}K_i}{|\omega|^2}\right)\right]|\phi_k|^2 
\nonumber \\
&\equiv& -\chi^i_{\phi} \frac{\partial V_{0\parallel i}}{\partial r}+\Pi^i_{res}, 
\\
\label{ion_Reynolds1}
\Pi^Z_{r||}=\langle\delta v_{E_r}\delta v_{\parallel Z}
\rangle &=& \left(\frac{c_s\rho_s}{L_{n_e}}\right)^2\sum_{\vec{k}} \frac{\gamma k_y^2}{|\omega|^2}\left[-\hat{V}_{0\parallel Z}'+\frac{Z}{\mu}\left< \frac{k_{\parallel}}{k_y} \right> \left(1+\frac{2k_y\omega_rL_{eZ}K_Z}{Z|\omega|^2}\right)\right]|\phi_k|^2\
\nonumber \\
&\equiv& -\chi^Z_{\phi}\frac{\partial V_{0\parallel Z}}{\partial r}+\Pi^Z_{res},
\label{imp_Reynolds1}
\end{eqnarray}
\noindent where
\begin{eqnarray}
\chi^i_{\phi} &=& \chi^Z_{\phi}=\left(\frac{c_s\rho_s}
{L_{n_e}}\right)^2\sum_{\vec{k}}\frac{\gamma k_y^2}{|\omega|^2}|\phi_k|^2, \\
\label{diffusivity}
\Pi^i_{res} &=& \left(\frac{c_s\rho_s}{L_{n_e}}\right)^2\sum_{\vec{k}}\gamma\frac{k_y \left< k_{\parallel}\right>}{|\omega|^2}\left(1+\frac{2k_y\omega_rL_{ei}K_i}{|\omega|^2}\right)|\phi_k|^2, \\
\label{residual_ion}
\Pi^Z_{res} &=& \left(\frac{c_s\rho_s}{L_{n_e}}\right)^2\sum_{\vec{k}}\gamma\frac{Z}{\mu}\frac{k_y \left< k_{\parallel} \right>}{|\omega|^2}\left(1+\frac{2k_y\omega_rL_{eZ}K_Z}{Z|\omega|^2}\right)|\phi_k|^2.
\label{residual_impurity}
\end{eqnarray}

\noindent Equations~(\ref{tot_Reynolds}) to (\ref{residual_impurity}) 
are main results of the present work. They describe 
$\Pi^{\rm eff}_{r\parallel}$ driven by ITG turbulence modified by the
presence of a single impurity species. 

The most important impurity
effects on $\Pi^{\rm eff}_{r\parallel}$ are implicitly
contained in 
Eqs.~(20), (\ref{diffusivity}), and 
(\ref{residual_impurity}). 
The characteristics of ITG turbulence, such as
the growth rate, the mode shift giving rise to $\left< k_\parallel \right>$, 
and the fluctuation amplitude, 
change as impurity characteristics vary, as shown in Sec.~III. 
Therefore, a complete study of impurity effects on $\Pi_{res}^{\rm eff}$ 
requires calculations of actual value of $|\phi_k|^2$ by the variation
of impurity characteristics. This is impossible in
the quasi-linear theory. 
To overcome
this limitation of the quasi-linear theory, we evaluate the ratio of
$\Pi^{\rm {eff}}_{res}$ to the diffusive flux [$\Pi^{\rm {eff}}_D = -\chi^{\rm {eff}}_{\phi} (\partial V_{0\parallel}/\partial r)$] 
for a fixed equilibrium flow gradient,
\begin{equation}
R = \frac{\Pi^{\rm {eff}}_{res}}{\Pi^{\rm {eff}}_{D} } = -\frac{\Pi^{\rm {eff}}_{res}}{\chi^{\rm {eff}}_{\phi} 
{\partial V_{0\parallel}}/{\partial r}}.
\label{R}
\end{equation}
\noindent Then, $R$ can be used as a quantitative measure, in the
context of the quasi-linear theory, of the relative
importance of the residual stress when impurity characteristics vary. 
Increase of $R$ implies the 
grow of the relative importance of $\Pi_{res}$ compared to 
$\Pi_D$, and vice versa.
Note that $R$ does not involve
$|\phi_k|^2$ term, which eliminates the difficulty of the quasi-linear theory
in predicting the level of turbulence amplitude. 
%
%
%
\begin{figure}
\begin{center}
\includegraphics[scale=0.7]{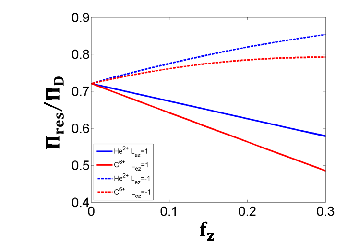}
\caption{The ratio of the residual stress ($\Pi_{res}$) to 
the diffusive momentum flux ($\Pi_D$) as a function
of $f_Z$. {$\rm C^{6+}$} (red) and {$\rm He^{2+}$} (blue) are
considered. Dotted and
solid lines represent the cases when $L_{eZ}=-1$ and 1, respectively.
The quasi-linear momentum flux is calculated by integrating over all $k_y$ 
values giving rise to instability. 
Other plasma parameters being used are same as those of Fig.~2.}
\label{ratio_fz}
\end{center}
\end{figure}

\begin{figure}
\includegraphics[scale=0.8]{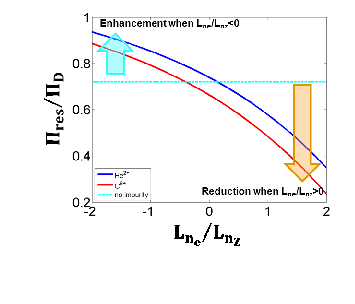}
\caption{The ratio of the residual stress ($\Pi_{res}$) to 
the diffusive momentum flux ($\Pi_D$) as a function
of $L_{eZ}=L_{n_e}/L_{n_Z}$. {$\rm C^{6+}$} (red) and {$\rm He^{2+}$} (blue) are
considered with a fixed value of $f_Z=0.3$. 
A cyan 
dotted line represents the $\Pi_{res}/\Pi_D$ value without impurities.
Other plasma parameters being used are same as those of Fig.~5.}
\label{ratio_Lez}
\end{figure}

Figure~\ref{ratio_fz} shows $R$ as a function of $f_Z$ for 
${\rm C}^{6+}$ (red) and ${\rm He}^{2+}$ (blue) impurities.
Plasma parameters being used to produce Fig.~\ref{ratio_fz} are
same as those of Fig.~2. In calculating the radial momentum flux in 
Figs.~5 and 6, we keep all the unstable modes 
in Eqs.~(20), (21), and (\ref{residual_impurity}) by integrating over 
all $k_y$ giving rise to positive growth rate.
%
%
%
%
$R$ shows a remarkably different
trend depending on the alignment of an impurity profile to 
the main ion density 
profile. It increases when two profiles have 
a different sign (dotted lines) while it decreases
when they have the same sign (solid lines).
For example, the reduction in $R$ as $f_Z$ changes
from $0$ to $0.15$ is $9.6\%$ and $16.6\%$ for $\rm{He}^{2+}$
and $\rm{C}^{6+}$ cases, respectively.
For a same absolute value of $L_{eZ}$, 
the decrease rate of $R$ when $L_{eZ}>0$ is larger
than the increase rate when $L_{eZ}<0$. 
Figure~\ref{ratio_Lez} highlights the impact of the profile alignment on
$R$ for a fixed value of $f_Z=0.3$. 
The cyan dotted line in Fig.~\ref{ratio_Lez}
represents the $R$ value when $f_Z=0$. It shows a considerable
reduction from its reference value (dotted line) as $L_{ne}/L_{nZ}$
increases. 

To understand the results 
in Figs.~\ref{ratio_fz} and \ref{ratio_Lez}, we evaluate
$R$ for a fixed equilibrium velocity gradient and a $k_y$ 
value corresponding to the most unstable mode. 
Using the quasi-neutrality condition, Eq.~(\ref{tot_residual}) 
becomes
\begin{eqnarray}
\Pi_{res}^{\rm {eff}} &=& \frac{1}{1+\hat{M}}\left(\frac{c_s\rho_s}{L_{n_e}}\right)^2\sum_{\vec{k}}\gamma\frac{k_yk_{\parallel}}{|\omega|^2}
\frac{1}{1-f_Z} \nonumber \\
&\times&\left\{1+\frac{2k_y\omega_r}{|\omega|^2}L_{ei}
\left[K_i-f_Z \left(K_i-\frac{L_{eZ}}{L_{ei}}\frac{K_Z}{Z} \right)\right]\right\}
|\phi_k|^2 ,
\end{eqnarray}
\noindent where the relation $\hat{M}={m_Zn_{Z0}}/{m_in_{i0}}=({\mu}/{Z})\left[{f_Z}/{1-f_Z}\right]$ 
is used. Next, noting that 
$\chi^{\rm {eff}}_\phi=(\chi^i_\phi+\hat{M}\chi^Z_\phi)/(1+\hat{M}) = \chi^i_\phi$
from Eq.~(\ref{diffusivity}), one can easily show that
\begin{equation}
\left< R\right>_{k_y} \sim \frac{1}{1+\hat{M}}
\left<\frac{k_\parallel}{k_y}\right>
\frac{1}{1-f_Z}
\left\{1+\frac{2k_y\omega_r}{|\omega|^2}L_{ei}
\left[K_i-f_Z \left(K_i-\frac{L_{eZ}}{L_{ei}}\frac{K_Z}{Z} \right)\right]\right\},
\label{R}
\end{equation}
\noindent where $\left< R\right>_{k_y}$ denotes an ensemble averaged value
of $R$ for a fixed $k_y$.

From the examination of Eq.~(\ref{R}), 
one can see that impurities affect $R$ through four ways: 
\begin{itemize}
\item[(1)] the mode shift which is contained in $\left<k_\parallel/k_y\right>$,
\item[(2)]impurity contents ($f_Z$),
\item[(3)] stability change by the change of the sign of $L_{eZ}$,
\item[(4)] the radial inhomogeneity of a impurity profile represented by
$K_Z$, {\it i.e.,} the direct impurity effect to the residual stress. 
\end{itemize}

\noindent 
All the first three effects turn out to be important in the determination of $R$.
Among them, the mode shift, which
is presented in Fig.~4, is found to be the most 
crucial factor determining the amount of residual stress by the variation 
of $L_{eZ}$.  
The last term in Eq.~(\ref{R}) represents a driving term to the
residual stress solely from the free energy of impurities, {\it i.e.,}
the radial inhomogeneity of an impurity profile.
Positive $L_{eZ}$ makes the last term of Eq.~(\ref{R}) negative
since $L_{ei} > 0$ in nominal tokamak discharges and $\omega_r <0$.
Thus, this term can further reduce $\left<R\right>_{k_y}$  combined with a strong reduction of $|X_0|$ when $L_{eZ} > 0$. However, the effect of
this term is marginal for cold impurity ions or high $Z$ impurities
like $W^{46+}$ expected in machines with full tungsten PFCs\cite{Angioni-JET}. 
However, this term may play
a role in $\left<R \right>$ when fusion-born 
helium ashes are present because of their small charge number and
natural tendency to align with main ions. 
Thus, results in this section 
might raise a potential concern to ITER operation where 
a significant amount of plasma rotation is envisioned to be produced
via intrinsic rotation.

\section{Summary and conclusions}\label{conclusion}

In conclusion, this paper has shown that impurities may change 
the amount of intrinsic rotation by affecting the residual stress
driven by turbulence.
To make an analytic study, we employed a two fluid model for 
ions and impurities for ITG turbulence in sheared slab geometry. 
The ion and impurity velocities are assumed to be equal, which is strictly
valid in the collisional regime. Then, we evaluated the 
momentum diffusivity and the residual stress in a coupled main ion-impurity
system. The principal findings of this paper are summarized as follows:
\begin{itemize}
\item[$\bullet$] Characteristics of ITG modes, such as
the growth rate or the mode structure, changes significantly 
by the variation of impurities contents
and/or the alignment of an impurity profile with that of main ions. 
Quasi-linear calculations show that
these changes result in a 
considerable change of turbulence driven residual stress, hence 
the amount of intrinsic rotation.

\item[$\bullet$] 
The ratio of residual stress ($\Pi_{res}$) to the diffusive momentum flux 
($\Pi_D$), $R=\Pi_{res}/\Pi_D$, 
can be used as a quantitative measure of relative importance of $\Pi_{res}$ 
over $\Pi_{D}$. 

\item[$\bullet$] For a core-peaked
impurity profile shown in Fig.~1(a) schematically, $R$ decreases
due to the stabilization and a large increase of the mode shift of the
unstable mode. For a wall-peaked 
impurity profile shown in Fig.~1(b) schematically, $R$ increases
a little implying the predominance of $\Pi_{res}$ over $\Pi_D$. 
The degree of the enhancement of $R$ when $L_{eZ}<0$ is smaller than that
of reduction when $L_{eZ}<0$ for a given impurity content. 
\end{itemize}

As a possible implication of the present study, we
pointed out that the amount of
intrinsic rotation in burning plasma experiments might be reduced.
This is due to the presence of  ${\rm He^{2+}}$ ashes  
which is likely to be aligned with the main ion profile. For example,
approximately $\sim 10\%$ of intrinsic rotation loss is expected if we
assume the presence of $\rm{He}^{2+}$ ions with $f_Z=0.15$ well-aligned
with a main ion profile. 
Such a well-aligned impurity profile is also possible in present
day devices, as shown in recent experiments in KSTAR\cite{KSTAR},
due to the prevalence of turbulent impurity transport. 
We note that this may cause the decoupling of 
the ion temperature and the toroidal rotation 
profiles, which has been observed in recent KSTAR H-mode experiments\cite{WKo}. 
When $L_{eZ}>0$, the reduction of residual stress 
will be larger than that of the diffusive momentum flux.
Since the intrinsic torque is believed
to be maximized near the pedestal top
region as shown in recent flux-driven gyrofluid
simulations\cite{JKPS,Tokunaga}, 
the reduction of the residual stress due to impurities will result in the 
reduction of a net intrinsic torque. This will reduce the total amount 
of toroidal rotation when an edge pedestal build up. 
To confirm this scenario, measurements of a rotation
profile for a varying impurity contents is necessary. This is left
as a future investigation. 

As an extension of the present study, we plan to perform
a gyrokinetic analysis including impurities. 
Nonlinear gyrokinetic
or gyrofluid simulations are also planned to elucidate the 
nonlinear physical process
of intrinsic rotation generation and 
saturation in the presence of impurities. These will be outstanding 
research subjects which will be published in the future.

\begin{acknowledgements}
\noindent This research was supported by 
the R$\&$D Program through National Fusion 
Research Institute (NFRI) funded by the Ministry of Science, ICT and Future Planning 
of the Republic of Korea (NFRI-EN1541-1), and by the
same Ministry under the ITER technology R$\&$D 
programme (IN1504-1).
\end{acknowledgements}

%
\end{document}